\documentclass[preprint,3p,12pt,numbers,sort&compress]{elsarticle}
\usepackage{amssymb}
\usepackage{amsthm}
\usepackage{amsmath}
\usepackage{placeins}
\usepackage{float}
\usepackage{subcaption}
\usepackage{graphicx, caption}
\usepackage{color,soul,xcolor}
\usepackage{stmaryrd}
\usepackage{relsize}
\usepackage[colorlinks=true, allcolors=blue]{hyperref}

\newcommand{\bvss}[1]{\textcolor{blue}{#1}}
\newcommand {\eqn}[1] {Eq.~(\ref{#1})}

\newcommand {\fig}[1] {Fig.~\ref{#1}}

\journal{arxiv}
\begin{document}
\begin{frontmatter}
%% Title, authors and addresses
%% use the tnoteref command within \title for footnotes;
%% use the tnotetext command for theassociated footnote;
%% use the fnref command within \author or \address for footnotes;
%% use the fntext command for theassociated footnote;
%% use the corref command within \author for corresponding author footnotes;
%% use the cortext command for theassociated footnote;
%% use the ead command for the email address,
%% and the form \ead[url] for the home page:
%% \title{Title\tnoteref{label1}}
%% \tnotetext[label1]{}
%% \author{Name\corref{cor1}\fnref{label2}}
%% \ead{email address}
%% \ead[url]{home page}
%% \fntext[label2]{}
%% \cortext[cor1]{}
%% \affiliation{organization={},
%%             addressline={},
%%             city={},
%%             postcode={},
%%             state={},
%%             country={}}
%% \fntext[label3]{}
\title{Physics--Informed Machine Learning and Uncertainty Quantification for Mechanics of Heterogeneous Materials}
%% use optional labels to link authors explicitly to addresses:
%% \author[label1,label2]{}
%% \affiliation[label1]{organization={},
%%             addressline={},
%%             city={},
%%             postcode={},
%%             state={},
%%             country={}}
%%
%% \affiliation[label2]{organization={},
%%             addressline={},
%%             city={},
%%             postcode={},
%%             state={},
%%             country={}}

\author[inst1]{B V S S Bharadwaja}
\author[inst2]{Mohammad Amin Nabian}
\author[inst3]{Bharatkumar Sharma}
\author[inst2]{Sanjay Choudhry}
\author[inst1]{Alankar Alankar\corref{cor3}}
\ead{alankar.alankar@iitb.ac.in}
\cortext[cor3]{Corresponding author}

\affiliation[inst1]{organization={Department of Mechanical Engineering},%Department and Organization
            addressline={Indian Institute of technology Bombay, Powai}, 
            city={Mumbai},
%            postcode={400076}, 
%            state={Maharashtra},
            country={India}}
\affiliation[inst2]{organization={Nvidia},
            city={Santa Clara},
            country={USA}}
\affiliation[inst3]{organization={Nvidia AI Technology Center},
            city={Bangalore},
            country={India}}

%\author[inst1,inst2]{Author Three}
% \affiliation[inst2]{organization={Department Two},%Department and Organization
%             addressline={Address Two}, 
%             city={City Two},
%             postcode={22222}, 
%             state={State Two},
%             country={Country Two}}
%
\begin{abstract}
In this work, a model based on the Physics-Informed Neural Networks (PINNs) for solving elastic deformation of heterogeneous solids and associated Uncertainty Quantification (UQ) is presented. For the present study, the PINNs framework - Modulus developed by Nvidia is utilized, wherein we implement a module for mechanics of heterogeneous solids. We use PINNs to approximate momentum balance by assuming isotropic linear elastic constitutive behavior against a loss function. Along with governing equations, the associated initial / boundary conditions also softly participate in the loss function. Solids where the heterogeneity manifests as voids (low elastic modulus regions) and fibers (high elastic modulus regions) in a matrix are analyzed, and the results are validated against solutions obtained from a commercial Finite Element (FE) analysis package. The present study also reveals that PINNs can capture the stress jumps precisely at the material interfaces. Additionally, the present study explores the advantages associated with the surrogate features in PINNs via the variation in geometry and material properties. The presented UQ studies suggest that the mean and standard deviation of the PINNs solution are in good agreement with Monte-Carlo FE results. The effective Young's modulus predicted by PINNs for single representative void and single fiber composites compare very well against the ones predicted by FE, which establishes the PINNs formulation as an efficient homogenization tool.
\end{abstract}
\begin{keyword}
%% keywords here, in the form: keyword \sep keyword
PINNs \sep heterogeneous materials \sep neural networks \sep solid mechanics
%% PACS codes here, in the form: \PACS code \sep code
%\PACS 0000 \sep 1111
%% MSC codes here, in the form: \MSC code \sep code
%% or \MSC[2008] code \sep code (2000 is the default)
%\MSC 0000 \sep 1111
\end{keyword}

\end{frontmatter}

%% \linenumbers

%% main text
\section{Introduction}
\label{sec:intro}
Fiber--reinforced composites (FRCs) are made up of a matrix structure with low modulus of elasticity embedded with fibers of very high modulus of elasticity. Voids may be introduced during the manufacturing process of composites. Fibers and voids may be introduced strategically in architectured materials for applications where functional properties are required e.g. lightweightness. Total volume fraction and spatial distribution of voids and fibers govern the overall stiffness of composites. Stress concentration and incompatibility at interfaces of voids / fibers and matrix may act as sources of cracks and other types of structural defects \cite{mehdikhani2019}. Thus, a critical assessment of the behavior of composites against loading conditions is required to estimate the robustness of a composite design. Numerical studies have helped understand the design and performance of composites \cite{huang2005effects,malakooti2013multi}.
\par Traditional physics--based models in science and engineering are limited by the current physical understanding of phenomena, approximations, and qualitative insights. These models may not capture the complete complexity of the science and engineering problem. In the context of developing predictive models for complex engineering applications, the computational cost of traditional physics--based models increases disproportionately when experimentally acceptable accuracy and a realistic size of simulation domain are required.
\par In keeping with the recent and extraordinary success of Machine Learning (ML) algorithms in applications such as Computer Vision (CV) and Natural Language Processing (NLP), machine learning--based models are playing an increasingly important role in core scientific research as an alternative to physics--based models. However, within ML approach, data--driven models \cite{beniwal2019deep,revi2021machine} are linked to another class of disadvantages. While these models provide lower--dimensional high throughput alternatives to traditional scientific and engineering models, they take the form of black--box algorithms in typical applications. Consequently, interpretability and system--specific mechanistic insights that characterize typical scientific explanations are lost.
\par Machine learning algorithms are further complicated by a wide range of possible features and hyperparameters. More fundamentally, an ill--informed application of machine learning algorithms to scientific problems often ignores critical physical constraints, such as conservation laws. The requirement of a large amount of data further limits the application of a data--driven ML model for a practical problem \cite{Goodfellow-et-al-2016,bhutada2021machine}.
\par These limitations drive the need for interpretable ML algorithms that are integrated by the robust underlying physics while being predictive. Most currently available ML approaches fall short of providing fully interpretable models. Integrated ML models with well--established physical laws are sought for reliable predictions to fill the gap. Current research in Artificial Intelligence (AI) for science and engineering is seeing an ever-increasing inclination towards physics--driven machine learning models that seek to blend the best of both approaches by eliminating their drawbacks.
\par One such class of models within the regime of AI and ML is Deep Learning (DL)--based on Physics--Informed Neural Networks (PINNs). PINNs can integrate data with mathematical frameworks of physical laws such that the loss functions are established with the help of Partial Differential Equations (PDEs) and Boundary Conditions (BCs). PINNs can be used to solve PDEs even with the minimal or complete absence of data \cite{lagaris1998artificial,raissi2017physics}. Lagaris et al. \cite{lagaris1998artificial} was the first to propose the solution of PDEs using Artificial Neural Networks (ANNs) in 1998. However, due to the lack of computational resources and efficient algorithms, this method could not attract much attention. PINNs are being used to handle a variety of boundary value problems that are highly nonlinear \cite{hornik1991approximation,cybenko1989approximation,rudy2019data,wight2020solving,nguyen2020deep}. In order to solve the nonlinear PDEs, apriori linearization of the governing equations is not necessary for PINNs. The non-linearity is introduced using transfer functions across various \emph{layers} of PINNs.
%Noteworthy is that the inherent nonlinear transfer functions \alr{new reader may not know what are transfer functions...somehow need to articulate that the functions that are used for transferring information across layers...but very concisely} approximate the non--linearity in the governing equation. This turns out to be the main advantage of PINNs \alr{why is this an advantage ?}.\bvss{Advantage is that apriori linearization of non linear governing equations is not necessary whereas in standard numerical techniques we need to linearize}
%
\par PINNs have proved their ability to solve a variety of time dependent \cite{raissi2017physics} and independent \cite{haghighat2021physics} Boundary Value Problems (BVPs). PINNs have already been extended to scientific fields such as fluid mechanics \cite{yang2019predictive,jin2021nsfnets,mahmoudabadbozchelou2022nn,almqvist2021fundamentals}, heat transfer \cite{cai2021physics,mishra2021physics,niaki2021physics}, multi--physics \cite{niaki2021physics} and multi--scale \cite{rocha2021deepbnd} problems. Raissi et al. \cite{raissi2017physics,raissi2017physicsII} proposed the PINNs as an alternative approach to solve governing equations. The authors employed the data--driven approach to find solution of highly nonlinear Burger's \cite{basdevant1986spectral}, Schr\"odinger \cite{schrodinger1926quantisierung} and Allen--Cahn \cite{allen1979microscopic} equations. PINNs framework has been used to model the forward and inverse problems in elasticity imaging\footnote{Elasticity imaging is an experimental technique where the responses of biological tissues are measured for various loadings.} for biological tissues subjected to tensile loading \cite{zhang2020physics}. Combining the collocation method and deep neural networks (DNNs), Abueidda et al. \cite{abueidda2021meshless} developed a Deep Collocation Method (DCM) which is meshfree and independent of data. The authors showed that DCM is capable of predicting the solutions for highly non--linear material models like hyperelasticity, elastoplasticity.
%%%
%
\par PINNs successfully solve fourth--order PDEs \cite{wight2020solving}, which often occur in engineering problems. Whereas, traditional solvers like FE find it challenging to solve the fourth--order PDEs because of the requirement of $C^1$ continuity \cite{kirby2019code}. In this context, PINNs successfully approximate the solutions for biharmonic equations \cite{vahab2021physics,guo2021deep} which arise in elasticity and application of Cahn--Hillard equations \cite{wight2020solving}.
\par One special feature of PINNs, when compared to standard numerical solvers, is surrogate modeling and uncertainty quantification (UQ). One needs to resort to numerical techniques such as Monte Carlo (MC) to find an approximate solution to a stochastic PDE. Whereas PINNs can capture the incertitude behaviour that arises in stochastic PDEs even in the absence of labeled data \cite{zhu2019physics}. The stochastic nature of material properties and design space can influence the response of a complex combination of structure and loading \cite{KUMAR2020112538}. UQ provides a statistical distribution of system response. Since PINNs are built upon a spectrum of inputs, they can inherently predict a distribution of corresponding outputs. Thus, PINNs can be efficiently used for UQ for a well--set problem. Examples of UQ with MC are found in the work of Butler et al. \cite{Butler_MC}. MC--based UQs are the simplest and the most reliable methods. However, their application in conjunction with PINNs is still rare \cite{nabian2018deep}.
\par In the present work, the application of PINNs has been demonstrated for the elastic deformation of heterogeneous solids. Three classes of heterogenous solids have been considered, i.e., solids with voids, solids with second phases with high elastic modulus i.e. fibers, and solids having both voids and second phases with high elastic modulus. FRCs are tailor--made materials that possess mutually exclusive properties like strength and toughness. In order to utilize the best performance of the FRCs, it is very critical to estimate their optimum properties. In this context, standard numerical approaches spanning microscale to macroscale have been attempted \cite{kanoute2009multiscale}.
\par The purpose of the present work is twofold -- (1) to showcase the ability of PINNs for the analyses of primitive heterogeneities like single fibers and voids, (2) UQ associated with material and geometry using surrogate modeling. A series of case studies involving various sizes of fibers, voids and their distributions under uniform tension and shear loading are conducted. PINNs are used for solving the PDE representing balance of momentum in conjunction with constitutive equations of elasticity. This is performed without aid of apriori data of material behavior. Material properties representing linear isotropic elasticity are considered. Tests are performed on combined voids and fibers that mimic the real composites in nature. Stress fields are extracted from PINNs and validated against a standard commercial FE software. Finally, we employ PINNs for solving random differential equations arising from random geometry and material properties. The obtained statistical measures - mean, standard deviation and probability density distribution from surrogate modeling using PINNs are compared against those from Monte--Carlo FE.
\par The manuscript is laid as following. Section \ref{sec:governing_equations} describes the governing equations. Section \ref{sec:nn_solution} describes the problem setup in the framework of neural networks. Section \ref{sec:numerical_examples} demonstrates the application of the present PINN--based models for elastic deformation of various heterogeneous solids. MC based UQ studies are discussed in Section \ref{sec:uq}, and concluding remarks are presented in Section \ref{sec:conclusions}.
\section{Governing equations}
\label{sec:governing_equations}
\begin{figure}[H]
\centering
     \includegraphics[width=0.4\textwidth]{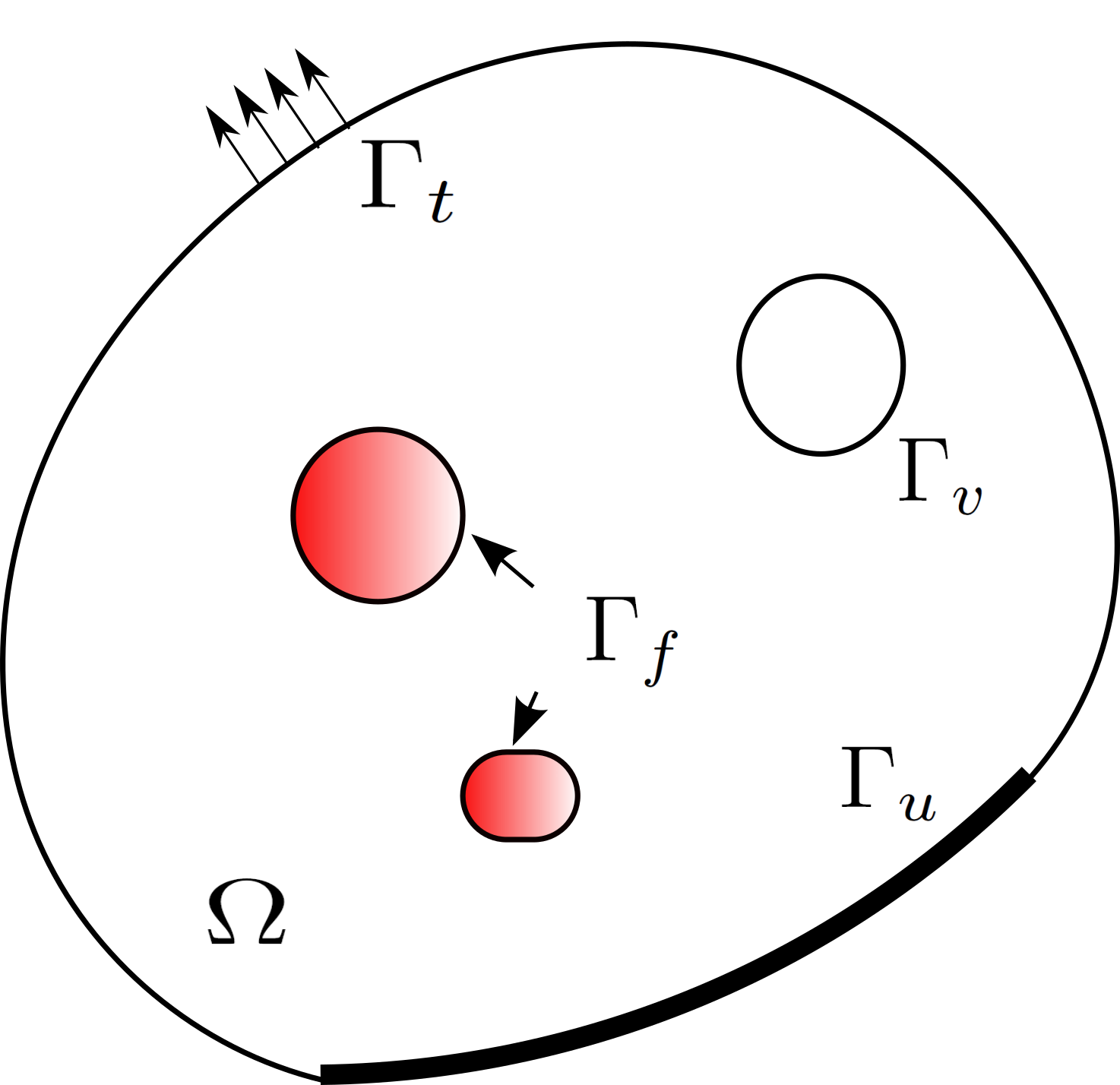}
     \caption{Definition of the domain showing boundaries of Dirchlet, Neumann, fibers and voids.}
     \label{fig:boundaries}
\end{figure}

A body represented by domain $\Omega\subset\mathbb{R}^2$ with boundary $\Gamma$ is shown in Figure \ref{fig:boundaries}. $\Gamma$ comprises of Dirichlet ($\Gamma_u$), Neumann ($\Gamma_t$), fibers ($\Gamma_f$) and voids ($\Gamma_v$). The boundaries of voids are maintained to be traction free. In case of fibers, the jump in tractions and displacements at interface are considered to be zero. i.e. the tractions and displacements are maintained to be continuous at the interface. The governing equations are given by
\begin{equation}
  {\nabla.{{\boldsymbol{\sigma}(\boldsymbol{x})}}} + \boldsymbol{b}(\boldsymbol{x}) = 0 \hspace{2mm} \boldsymbol{x}\in\Omega, 
\end{equation}
\noindent where $\nabla$ is the gradient operator,  $\textbf{u},\boldsymbol{\sigma},\boldsymbol{\epsilon}$ represent displacement vector, Cauchy stress tensor and strain tensor respectively. Body force $\boldsymbol{b}$ is assumed to be zero. The essential and natural boundary conditions are given by
\begin{subequations}
\begin{align}
\centering
\label{eq:bc_displacement}
  \boldsymbol{u}(\boldsymbol{x}) &= {\boldsymbol{u}^*}, \hspace{5mm} \Gamma=\Gamma_u,\\
\label{eq:bc_traction}
  {\boldsymbol{\sigma}(\boldsymbol{x})}.\boldsymbol{n} &= \boldsymbol{t}, \hspace{5mm} \Gamma=\Gamma_t, \\
\label{eq:bc_void}
  {\boldsymbol{\sigma}(\boldsymbol{x})}.\boldsymbol{n} &= 0., \hspace{5mm} \Gamma=\Gamma_v, \\
\label{eq:bc_inclusion}
  \llbracket{\boldsymbol{u}(\boldsymbol{x})}\rrbracket=0, \llbracket{\boldsymbol{\sigma}(\boldsymbol{x})}.\boldsymbol{n}\rrbracket &=0., \hspace{5mm} \Gamma=\Gamma_f,
\end{align}
\end{subequations}
where $\boldsymbol{n}$ is the surface normal, $\boldsymbol{t}$ is traction and $\boldsymbol{u}^*$ is the prescribed displacement.
For a linearly elastic, isotropic and homogeneous material the constitutive relations can be expressed as
\begin{equation}
 \boldsymbol{\sigma}(\boldsymbol{x}) = \boldsymbol{\mathcal{C}}:\boldsymbol{\epsilon}(\boldsymbol{x}),
\end{equation}
where $\boldsymbol{\mathcal{C}}$ is a fourth--rank elasticity tensor representing the relation between stress and strain tensors. Under the assumption of small strain theory, strain is represented by
\begin{equation}
  \boldsymbol{\epsilon} = \frac{1}{2}\left(\nabla {\boldsymbol{u}} + \nabla {\boldsymbol{u}}^T\right)\bvss{.}
\end{equation}
The numerical simulations in this paper are approximated under plane strain setup and the elasticity tensor $\boldsymbol{\mathcal{C}}$ in Voigt notation for plane strain condition is written as
\[
\begin{bmatrix}
    \lambda+2\mu & \lambda & 0 \\
    \lambda & \lambda+2\mu & 0 \\
    0 & 0 & \mu
\end{bmatrix}
.\]
where $\lambda$ and $\mu$ are the Lame's constants. The case studies presented in the following sections are considered under static case, i.e., time dependence is not considered as an input to the neural network. Non--dimensionalization of the governing equations helps with the optimizers to rapidly reduce the loss associated with the problem. Therefore, in the present study the stresses are non--dimensionalized with shear modulus $\mu$ and the spatial dimensions with the length of the domain $l$. The non--dimensionalized equations under static case are
\begin{equation}
  \tilde{\nabla}.{{\boldsymbol{\tilde{\sigma}}(\boldsymbol{\tilde{x}})}} = 0 \hspace{2mm} \boldsymbol{\tilde{x}}\in\Omega,
\end{equation}
where $\boldsymbol{\tilde{\sigma}}={\boldsymbol{\sigma}}{/ \mu}, \boldsymbol{\tilde{x}}={\boldsymbol{x}}{/l}$. Hence in the following sections all the stress and spatial quantities are non--dimensional and tilde exponent  will be dropped unless explicitly stated.

%\begin{equation}
%    div(\sigma)=\pho \frac{\partial u}{\partial t}
%\end{equation}
%

\section{A Neural Network Based Solution}
\label{sec:nn_solution}
The central idea of PINNs is to minimize a loss function that can be set up in strong form (differential form) or weak form (variational form) along with the given set of boundary and initial conditions. The minimization process involves calculating the loss function at certain collocation points at the interior of the domain and at the boundaries under consideration. Thus, PINNs can be classified as meshless methods which is another advantage when compared to standard numerical techniques. In case of FE, the meshes need to conform to the geometry of the fiber / void and are usually required to be very fine near interfaces. If appropriate mesh refinement is not adopted, there is a possibility of severe distortion and the solution may not be accurate. PINNs as meshfree method do not face such issue. Nevertheless, in case of PINNs the accuracy of the solution does depend on the number of collocation points.
\par
\begin{figure}[H]
\centering
     \includegraphics[width=1.0\textwidth]{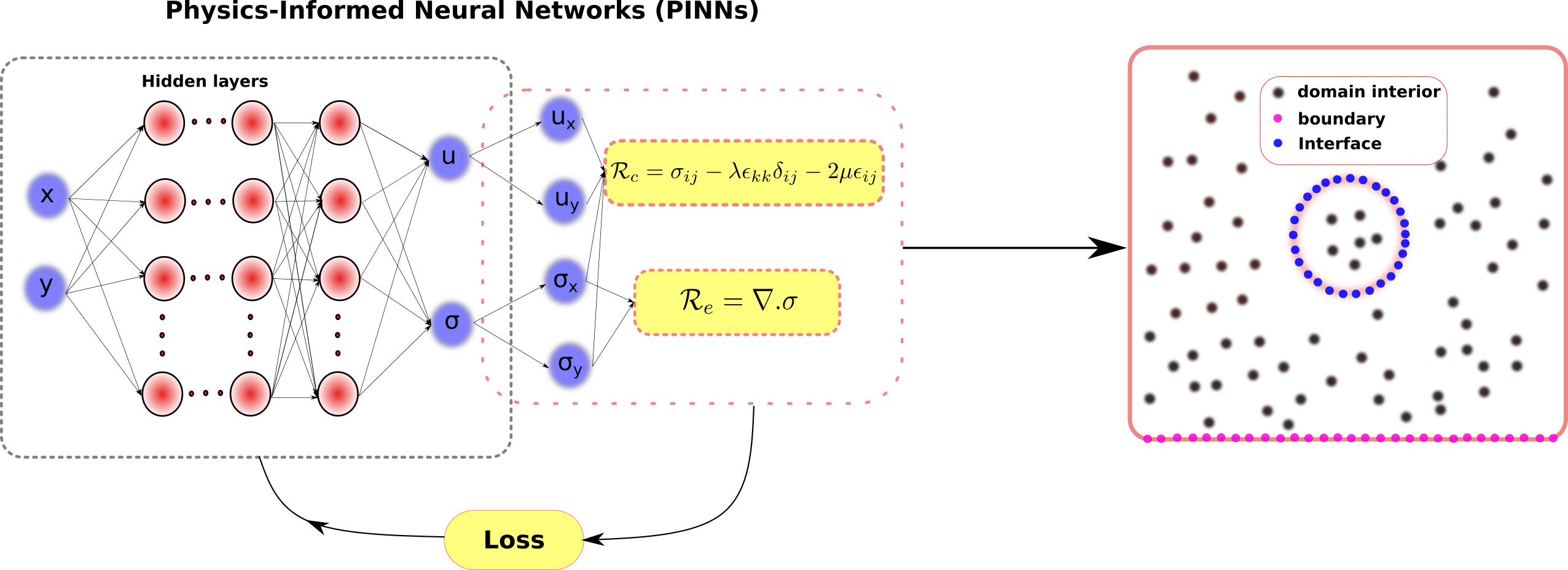}
     \caption{PINNs for time independent solid mechanics. On the right side is shown the distribution of batch points on various regions of a single fiber / void, domain interior, boundary and interface. No batch point is considered inside a void. $\mathcal{R}_e$ and $\mathcal{R}_c$ are the residues corresponding to linear momentum and constitutive equations that contribute to the total loss function.}
     \label{fig:Feed_Forward_neural_network}
\end{figure}

 A schematic of a PINN is shown in Figure \ref{fig:Feed_Forward_neural_network}. The network consists of spatial variables ($x,y$) as inputs and displacements ($u,v$) and stresses ($\sigma_{xx},\sigma_{yy},\sigma_{xy}$) as outputs of the NN. Using automatic differentiation, the residual arising from satisfying the governing and constitutive equations is minimized using a suitable optimizer. To find global minima of the proposed loss function, optimizers such as Adam \cite{kingma2014adam}, L--BFGS \cite{liu1989limited}, etc., are generally used in practice. The optimizer searches for the optimal weights and biases of the network which leads to the approximate solution of the governing equation. In the present work, Adam optimizer with a decaying learning rate algorithm is used.
\par The non--convex loss function is shown in \eqn{Eq.loss function} where $\mathcal{L}_r$ represents the total loss combining interior of the domain ($\Omega$) and its boundary conditions:
\begin{equation}
    \mathcal{L}_r = \lambda_\Omega\mathcal{L}_\Omega+\lambda_u\mathcal{L}_{\Gamma_u}+\lambda_t\mathcal{L}_{\Gamma_t}+\lambda_i\mathcal{L}_{\Gamma_f}+\lambda_v\mathcal{L}_{\Gamma_v},
    \label{Eq.loss function}
\end{equation}

\noindent where the parameters $\lambda_{\Gamma_{\Omega/u/t/f/v}}$ represent the weights associated with the interior domain and boundary conditions.
%\begin{equation}
\begin{subequations}
\label{eq:sublossfunctions}
\begin{align}
%\centering
\mathcal{L}_{\Omega} = \int_{\Omega} (\boldsymbol{\nabla}.\boldsymbol{\sigma})^2 \,d\Omega + \int_{\Omega} (\boldsymbol{\mathcal{C}:\epsilon-\sigma})^2 \, d\Omega,\\
%\label{eq_loss_governing_equation}
\mathcal{L}_{\Gamma_u}=\int_{\Gamma_u} (\boldsymbol{u-u^*})^2\,d\Gamma_u,\\
%\label{eq_loss_dirichlet}
\mathcal{L}_{\Gamma_t}=\int_{\Gamma_t} (\boldsymbol{\sigma.n}-\boldsymbol{t})^2\,d\Gamma_t, \\
%\label{eq_loss_traction}
\mathcal{L}_{\Gamma_f}=\int_{\Gamma_f} (\boldsymbol{u_m-u_f})^2 d\Gamma_f + \int_{\Gamma_f} (\boldsymbol{\sigma_m.n_m-\sigma_f.n_f})^2 d\Gamma_f, \\
\mathcal{L}_{\Gamma_v}=\int_{\Gamma_v} (\boldsymbol{\sigma_v.n_v-t_v})^2 d\Gamma_v,
\end{align}
\end{subequations}
where subscripts $m,f,v$ represent matrix, fiber and void, respectively. The integral form of loss function for the current problem is shown in \eqn{eq:sublossfunctions}. A Monte carlo approximation of this integral loss is considered using collocation points inside the domain. As shown in the above adopted loss function structure, there are many parameters that control the solution accuracy for a given problem. One can adjust the weights associated with the boundary conditions that will enforce them in a strong / weak manner.
\par Sampling points across the domain can be chosen in a variety of ways, i.e., uniform sampling, quasirandom sampling \cite{halton1960efficiency}, and importance sampling \cite{nabian2021efficient}. In uniform sampling, the points are generated by sampling from a uniform distribution. In contrast, in quasirandom sampling, the points are generated based on the low discrepancy sequences like Halton or Sobol sequences \cite{halton1960efficiency}. Based on the requirements, the number of collocation points are allocated differently on the boundaries and interior of the domain. In the present work, approximately 25,000 batch points in matrix, fiber and interface, each, with uniform sampling distribution are considered, whereas inside the void no batch points are considered. The maximum number of batch points is restricted by the GPU capacity and the number of GPUs.
\par To effectively capture the physics of the governing equations by improving the accuracy and convergence behavior of the solution, various DNN architectures have been proposed in the literature. Among those are -- Fourier networks where the activation functions are seen in the realm of Fourier space \cite{rahaman2019spectral,wang2021understanding}, Highway Fourier network \cite{huang2016deep} and Sinusoidal Representation Networks (SiReNs) \cite{sitzmann2020implicit}. Since Fourier networks have proven their ability in mitigating spectral bias  \cite{wang2021understanding}, the same was adopted for the present study. The network architecture chosen for the present study is a Fourier network, and the number of hidden layers is 6, with 512 neurons in each layer \cite{hennigh2021nvidia}.
\par In order to capture the non--linearity in the governing equation, the swish activation function \cite{ramachandran2017searching}, as given by the following equation, is employed:
\begin{equation}
    s(x)=\frac{x}{1+e^{-\beta x}}.
    \label{Eq.swish function}
\end{equation}
Varying the parameter $\beta$ in \eqn{Eq.swish function} gives a wide spectrum of non--linear functions where $\beta\to0$ makes a scaled linear function, whereas $\beta\to\infty$ manifests the Rectified Linear Unit (ReLU) \cite{klambauer2017self} function. Sigmoid--weighted Linear Unit (SiLU) \cite{elfwing2018sigmoid} is a special form of swish with $\beta=1$ is chosen for the present study.
\par In the present study, the Modulus framework \cite{hennigh2021nvidia} developed by NVIDIA is adopted in order to solve the governing equations of the system under consideration. Modulus is based on the concept of PINNs where the solution of a PDE is approximated as $\Tilde{u}=\mathcal{N}(x,y/\mathbf{W,b})$ where $\mathcal{N}$ represents a non--linear map between input variables $(x,y)$ and output variables $(\boldsymbol{u},\boldsymbol{\sigma})$, with $(\mathbf{W,b})$ as the unknown parameters that can be found via minimization of the loss function. Vectors $\mathbf{W}$ and $\mathbf{b}$ contain weights and biases \cite{Goodfellow-et-al-2016} associated with various neurons across the layers of PINNs.
\par The most efficient way of constructing the neural network, especially for solid mechanics, is to have two separate networks for displacements and stresses \cite{rao2021physics}. The same methodology is adopted while constructing the neural network for all the problems under consideration.
\par The hardware used for all the representative numerical simulations in the present study is one Tesla V100 GPU card along with 64 cores of Intel Xeon Silver CPUs and 64 GB of RAM. The effective number of batch points can also be increased by increasing the number of GPUs.
\section{Numerical examples}
\label{sec:numerical_examples}
%\hl {Should we show one example of homogeneous domain ?}.
\par
\begin{figure}[H]
 \centering
     \includegraphics[width=0.4\textwidth]{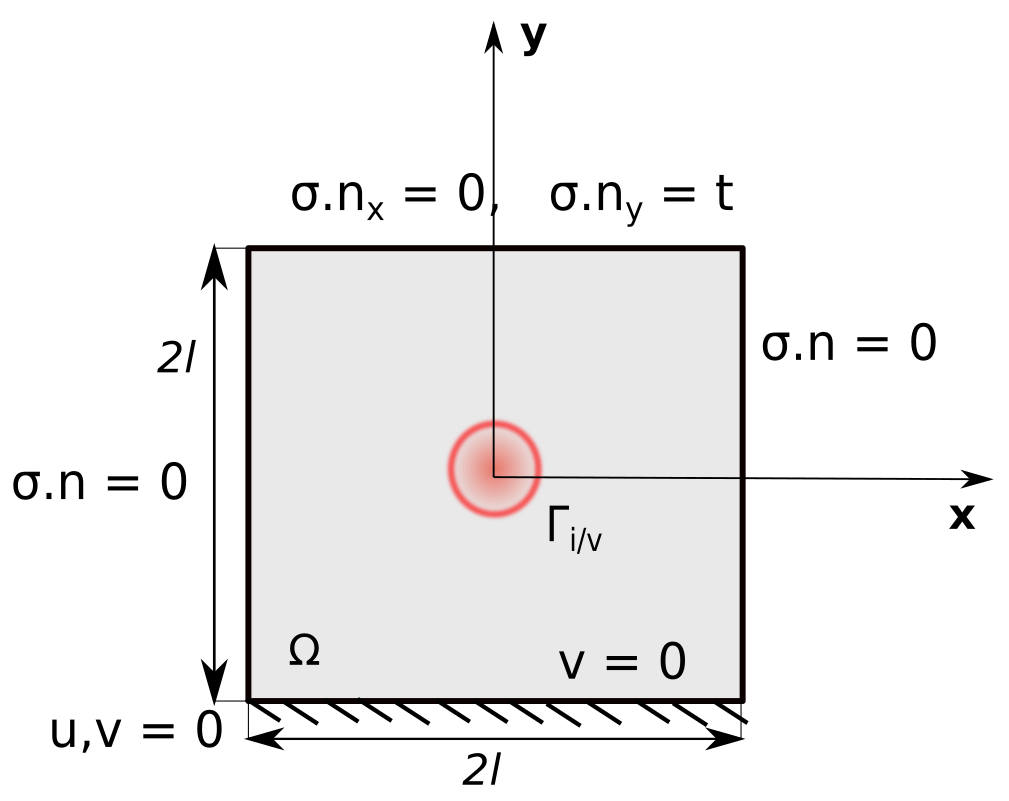}
     \caption{Initial setup of the problem under uniaxial tension.}
     \label{fig:Initial_setup_of_the_problem}
\end{figure}

 In this section, various case studies are presented to showcase the ability of PINNs to capture highly discontinuous stress fields at the interface of fibers and voids.
%Results from the current PINNs approach show an excellent comparison against the one from a standard Galerkin FE approach.
An example of the initial setup of the problem under uniaxial tension is shown in Figure. \ref{fig:Initial_setup_of_the_problem}.
Table \ref{table:parameters list} represents a list of parameters chosen for the following numerical simulations where $E_f$ represents Young's modulus of the fiber, and $E_m$ represents Young's modulus of the matrix. The ratio of Young's modulus for fibers is five times that of the matrix for all the simulations.
\begin{table}[h!]
\centering
\caption{Parameters chosen for the numerical studies.}
\label{table:parameters list}
\begin{tabular}{c c} 
 \hline
 Parameter & Value\\
 \hline
  $E_f$ (GPa) & 200 \\ [0.5ex] 
 $E_m$ (GPa) & 40 \\
  Poisson's ratio $\nu$ & 0.25  \\ [1ex] 
 Half edge length $l$ (mm) & 1 \\
 \hline
\end{tabular}
\end{table}
\subsection{Single fibers and voids under uniaxial tension}
\label{sec:single_fibers}
The fibers are assumed to be perfectly bonded (no relative displacement of nodes at the interface), whereas voids are modeled as geometric discontinuities containing no matrix material inside them. The geometry module of Modulus is versatile enough such that one can generate the necessary geometry under consideration using Modulus itself. It contains all primitive shapes for both 2D (line, circle, triangle, rectangle) and 3D (sphere, cube). First, the single fiber and void with different shapes are studied i.e., circular, elliptical, and square. Tensile traction of magnitude 80 MPa is applied on the top surface of the domain, and the lateral faces are left traction--free. Detailed boundary conditions are shown in Eqs. (\ref{eq:bc_displacement})-(\ref{eq:bc_inclusion}).
\begin{figure}[H]%[!htbp]
\centering
     \includegraphics[width=0.45\textwidth]{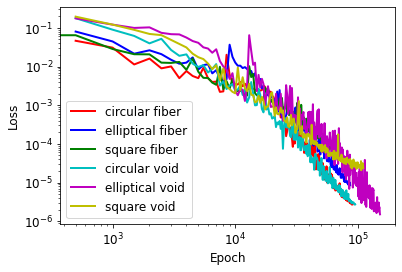}
     \caption{Convergence of loss function for single voids and fibers.}
  \label{fig:Convergence_of_loss_function_for_single_voids/inclusions}
\end{figure}

\begin{figure}[H]%[!htbp]
\centering
  \begin{subfigure}[H]{0.32\textwidth}
     \includegraphics[width=\textwidth]{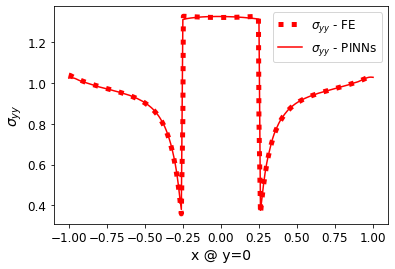}
     \caption{$\sigma_{yy}$}
     \label{fig:1}
  \end{subfigure}
  \begin{subfigure}[H]{0.32\textwidth}
     \includegraphics[width=\textwidth]{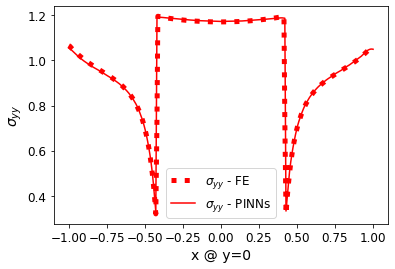}
     \caption{$\sigma_{yy}$}
     \label{fig:2}
  \end{subfigure}
  \begin{subfigure}[H]{0.32\textwidth}
     \includegraphics[width=\textwidth]{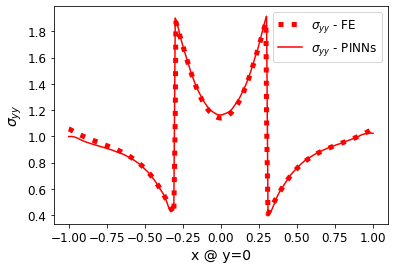}
     \caption{$\sigma_{yy}$}
     \label{fig:2}
  \end{subfigure}
  \caption{Comparison of PINNs against the solution obtained from commercial FE package for circular, elliptical and square fibers.}
  \label{fig:sigma_yy_comparison_line_plots_for_single_inclusions}
\end{figure}

\begin{figure}[H]
\centering
  \begin{subfigure}[!htbp]{0.32\textwidth}
     \includegraphics[width=\textwidth]{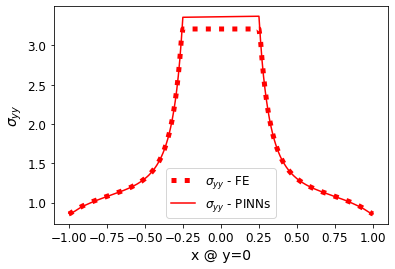}
     \caption{$\sigma_{xx}$}
     \label{fig:1}
  \end{subfigure}
  \begin{subfigure}[H]{0.32\textwidth}
     \includegraphics[width=\textwidth]{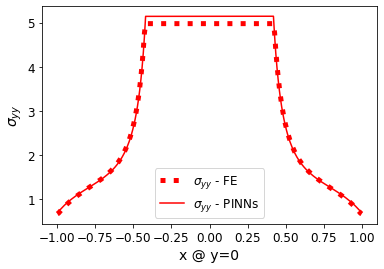}
     \caption{$\sigma_{yy}$}
     \label{fig:2}
  \end{subfigure}
  \begin{subfigure}[H]{0.32\textwidth}
     \includegraphics[width=\textwidth]{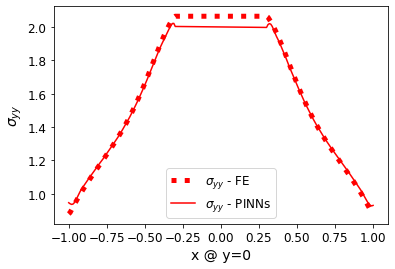}
     \caption{$\sigma_{yy}$}
     \label{fig:2}
  \end{subfigure}
  \caption{Comparison of PINNs against the solution obtained from commercial FE package for circular, elliptical and square voids.}
  \label{fig:sigma_yy_comparison_line_plots_for_single_voids}
\end{figure}
\par The domain under consideration is a square shape with various fibers / voids embedded inside the matrix at the center position, in each case. The radius of the circular fiber is 0.25 (non-dimensional). In the case of an elliptical fiber, the major / minor axis ratio is maintained at 2 with oblate in orientation.
\par Figure. \ref{fig:Convergence_of_loss_function_for_single_voids/inclusions} shows the degradation of loss over a number of epochs in logarithmic scale for different fibers / voids. Stress fields obtained from PINNs are extracted and compared against the FE solutions. Figures. \ref{fig:sigma_yy_comparison_line_plots_for_single_inclusions} and \ref{fig:sigma_yy_comparison_line_plots_for_single_voids} show the comparison of stress fields $\sigma_{yy}$ at $y=0$ for fibers and voids respectively.
\begin{figure}[H]%[!htbp]
\centering
  \begin{subfigure}[H]{0.48\textwidth}
     \includegraphics[width=\textwidth]{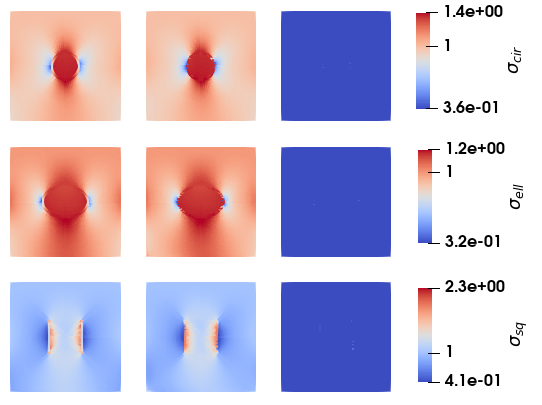}
     \caption{}
     \label{fig:1}
  \end{subfigure}
  \begin{subfigure}[H]{0.48\textwidth}
     \includegraphics[width=\textwidth]{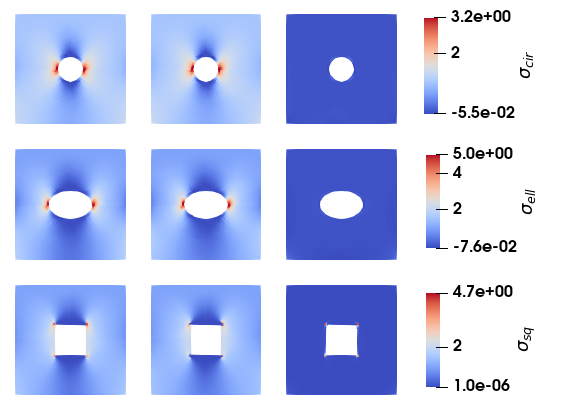}
     \caption{}
     \label{fig:2}
  \end{subfigure}
  \caption{$\sigma_{yy}$ contours for various (a) fibers and (b) voids. Each row contains FE, PINNs and their difference, in order.}
  \label{fig:contours_for_various_inclusions/voids.Each_row_contains_FE_Modulus_and_their_difference_in_order}
\end{figure}
\begin{figure}[H]
\centering
  \begin{subfigure}[H]{0.43\textwidth}
     \includegraphics[width=\textwidth]{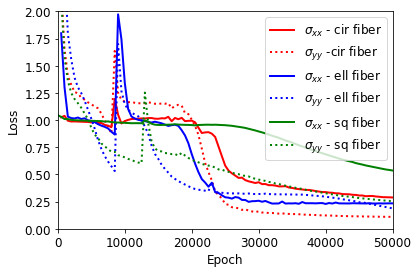}
     \caption{}
     \label{fig:1}
  \end{subfigure}
  \begin{subfigure}[H]{0.43\textwidth}
     \includegraphics[width=\textwidth]{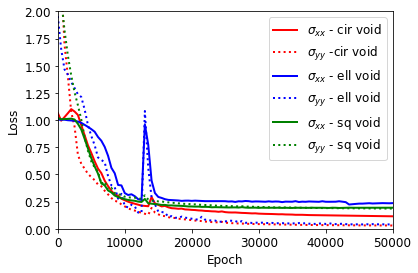}
     \caption{}
     \label{fig:2}
  \end{subfigure}
  \caption{Validation losses for single (a) fiber (b) void.}
  \label{fig:incls/voids_val_losses}
\end{figure}
\par The $\sigma_{yy}$ stress distribution for various fibers and voids obtained from PINNs are shown in Figure. \ref{fig:contours_for_various_inclusions/voids.Each_row_contains_FE_Modulus_and_their_difference_in_order}. Figure. \ref{fig:incls/voids_val_losses} shows the validation losses corresponding to stresses for all shapes of fibers and voids. After some epochs, all the losses get saturated, which indicates that the optimal solution is obtained with the chosen hyperparameters.
\subsection{Fiber reinforced composites subjected to tensile and shear loads}
In Section \ref{sec:single_fibers}, it is established that PINNs framework is able to capture the severe stress discontinuities when single voids and fibers are present in the domain. The real multi--phase structures are not limited to only a single fiber / void. They manifest in a large number, random distributions, and shapes. Thus, in this section, the ability of the developed model to predict stress fields precisely for FRCs where multiple fibers and void are arranged at a finite distance from each other, is presented.
%In the following simulations, Young's modulus ratio between the fiber and matrix is maintained at 5 with a Poisson's ratio of 0.25.
%
\subsubsection{FRCs with a couple of fibers or voids}
\begin{figure}[H]%[!htbp]
\centering
     \includegraphics[width=0.47\textwidth]{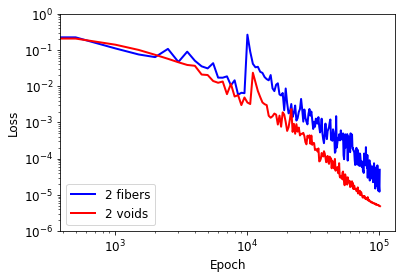}
     \caption{Convergence of loss function for two fibers and voids.}
  \label{fig:Adam_optimizer_loss_curves_for_couple_of_inclusions_and_voids}
\end{figure}

In this case study the fibers and voids are located at a finite distance of 0.2. The radius of fiber and voids is considered to be 0.25, and the domain length is $2l$ where $l$ is the half--edge length as shown in Figure. \ref{fig:Initial_setup_of_the_problem}. Tensile loading is applied and the boundary and interface conditions are the same as in the single fiber problem in Section. \ref{sec:single_fibers}. The loss curves in logarithmic scale corresponding to the solution are shown in Figure. \ref{fig:Adam_optimizer_loss_curves_for_couple_of_inclusions_and_voids}.
\begin{figure}[H]
\centering
  \begin{subfigure}[H]{0.4\textwidth}
     \includegraphics[width=\textwidth]{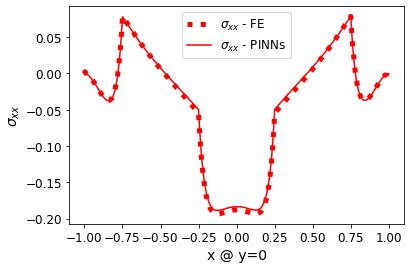}
     \caption{}
     \label{fig:1}
  \end{subfigure}
  \begin{subfigure}[H]{0.4\textwidth}
     \includegraphics[width=\textwidth]{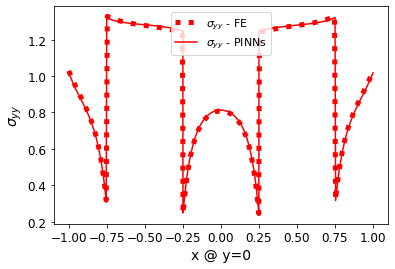}
     \caption{}
     \label{fig:2}
  \end{subfigure}
  \caption{Comparison of stress field for two fibers in the domain under simple tension. The domain is shown in \fig{fig:stress_contours_in_2_fibers/voids}. (a) $\sigma_{xx}$, (b) $\sigma_{yy}$.}
  \label{fig: 2 inclusions line plots}
\end{figure}
\begin{figure}[H]
\centering
  \begin{subfigure}[H]{0.4\textwidth}
     \includegraphics[width=\textwidth]{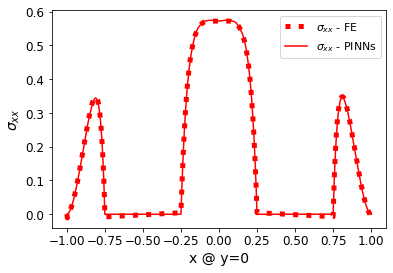}
     \caption{}
     \label{fig:1}
  \end{subfigure}
  \begin{subfigure}[H]{0.4\textwidth}
     \includegraphics[width=\textwidth]{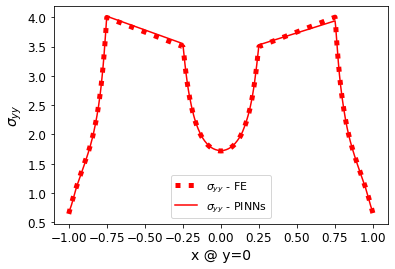}
     \caption{}
     \label{fig:2}
  \end{subfigure}
  \caption{Comparison of stress field for two voids in the domain under simple tension. The domain is shown in \fig{fig:stress_contours_in_2_fibers/voids}. (a) $\sigma_{xx}$, (b) $\sigma_{yy}$.}
  \label{fig: 2 voids line plots}
\end{figure}
\begin{figure}[H]%[!htbp]
\centering
  \begin{subfigure}[H]{0.48\textwidth}
     \includegraphics[width=\textwidth]{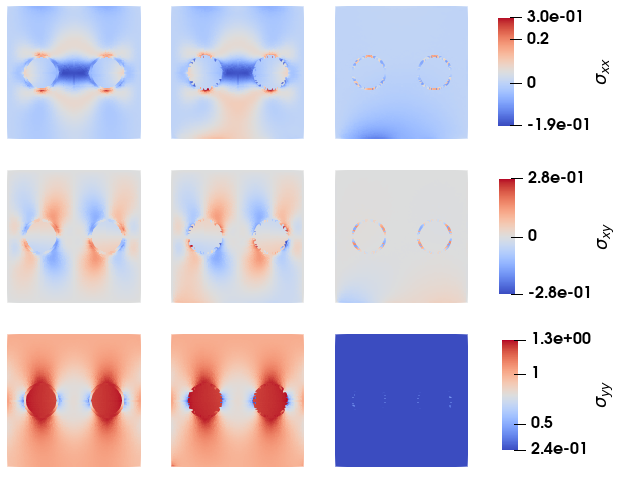}
     \caption{Tension}
     \label{fig:1}
  \end{subfigure}
  \begin{subfigure}[H]{0.48\textwidth}
     \includegraphics[width=\textwidth]{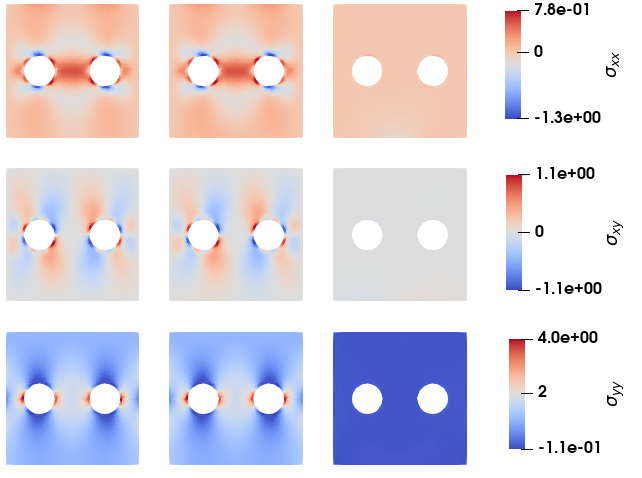}
     \caption{Shear}
     \label{fig:2}
  \end{subfigure}
  \caption{Comparison of stress distribution from PINNs against the solution obtained from commercial FE package for composites under tensile loading. Each row contains FE, PINNs and their difference in order.}
  \label{fig:stress_contours_in_2_fibers/voids}
\end{figure}
Figures. \ref{fig: 2 inclusions line plots} and \ref{fig: 2 voids line plots} show an excellent match for the stresses in the '$y$' direction from PINNs and FE. Stress contours and the difference between FE and PINNs solutions for $\sigma_{xx},\sigma_{xy},\sigma_{yy}$ are shown in Figure. \ref{fig:stress_contours_in_2_fibers/voids} and show a good agreement.
\subsubsection{FRCs with multiple fibers and voids}
In the present section, the composites are constructed by voids and elastic fibers placed at regular intervals in the matrix. In order to understand the actual behavior of composites, it is necessary to understand the interaction between voids and fibers. The domain under consideration has equally distributed circular shape voids and fibers. These composites are subjected to two different types of loading - tension and shear. The loads are applied on the top surface of magnitude 80 MPa.
\begin{figure}[H]%[!htbp]
\centering
     \includegraphics[width=0.5\textwidth]{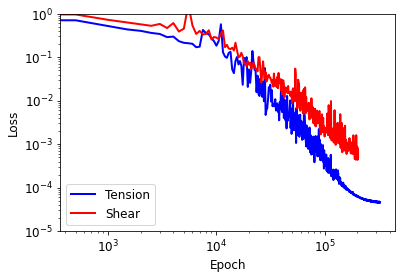}
     \caption{Convergence of loss function for FRCs subjected to tensile and shear loads.}
  \label{fig:Convergence_of_loss_function_for_composites}
\end{figure}
\par
\begin{figure}[H]
\centering
  \begin{subfigure}[H]{0.4\textwidth}
     \includegraphics[width=\textwidth]{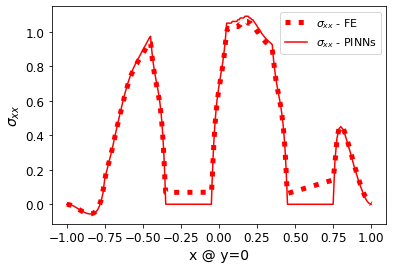}
     \caption{}
     \label{fig:1}
  \end{subfigure}
  \begin{subfigure}[H]{0.4\textwidth}
     \includegraphics[width=\textwidth]{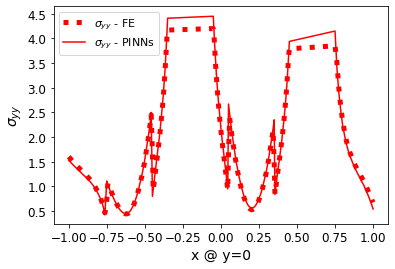}
     \caption{}
     \label{fig:2}
  \end{subfigure}
  \caption{Response of FRCs on application of tensile load. (a) $\sigma_{xx}$, (b) $\sigma_{yy}$.}
  \label{fig:Composites_under_tension}
\end{figure}

\begin{figure}[H]
\centering
  \begin{subfigure}[H]{0.4\textwidth}
     \includegraphics[width=\textwidth]{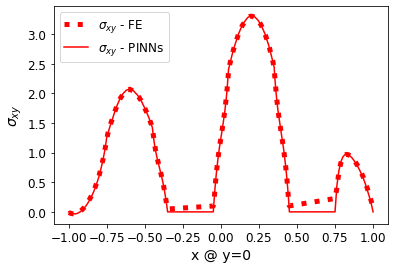}
     \caption{}
     \label{fig:1}
  \end{subfigure}
  \begin{subfigure}[H]{0.4\textwidth}
     \includegraphics[width=\textwidth]{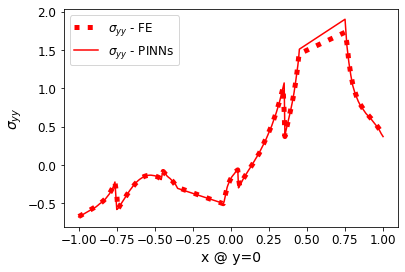}
     \caption{}
     \label{fig:2}
  \end{subfigure}
  \caption{Response of FRCs on application of shear load. (a) $\sigma_{xy}$, (b) $\sigma_{yy}$.}
  \label{fig:Composites_under_shear}
\end{figure}

\begin{figure}[H]%[!htbp]
\centering
  \begin{subfigure}[H]{0.48\textwidth}
     \includegraphics[width=\textwidth]{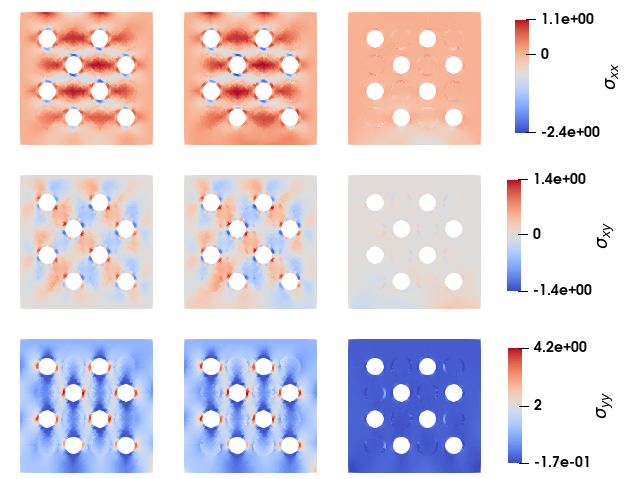}
     \caption{}
     \label{fig:1}
  \end{subfigure}
  \begin{subfigure}[H]{0.48\textwidth}
     \includegraphics[width=\textwidth]{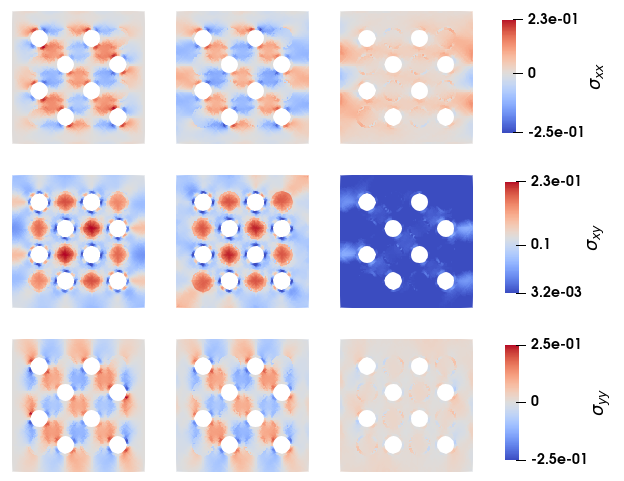}
     \caption{}
     \label{fig:2}
  \end{subfigure}
  \caption{Comparison of stress distribution from PINNs against the solution obtained from commercial FE package for tensile loading. Each row contains FE, PINNs and their difference, respectively: (a) Tension, (b) Shear.}
  \label{fig:Composites_stress_contours}
\end{figure}

 The convergence of loss function corresponding to tensile and shear loads on the composites in logarithmic scale is shown in the Figure. \ref{fig:Convergence_of_loss_function_for_composites}. The comparison of stress field predicted by PINNs as compared against the one obtained from FE at $y=0.2$ is shown in the Figures. \ref{fig:Composites_under_tension} - \ref{fig:Composites_under_shear}. Various stress contours corresponding to tension and shear are shown in Figure. \ref{fig:Composites_stress_contours}.

\subsection{Effect of high Young's modulus ratio}
\par
\begin{figure}[H]%[!htbp]
\centering
     \includegraphics[width=0.7\textwidth]{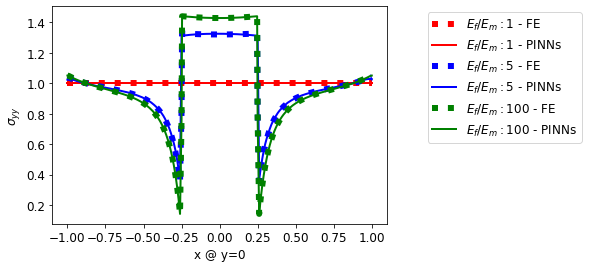}
     \caption{Comparison of PINNs based solution against the solution obtained from commercial FE package for various ratios of Young's modulus for matrix and fiber.}
     \label{fig:high_modulus_ratio}
\end{figure}

 The behavior of a composite can be altered by infusing soft and hard fibers of various sizes, and shapes \cite{falzone2016influences}. However, soft fibers alone degrade the effective material modulus of the composite, which in turn reduces the load carrying capacity of the composite \cite{peng2015situ}. However, if the soft fibers are accompanied by high modulus fibers strategically, composites will gain structural integrity \cite{fincato2019influence}. In the present section, the ability of PINNs to predict the stress discontinuities for a higher modulus ratio, is explored. Separate simulations for elasticity modulus ratios $E_f/E_m$ of 1, 5, and 100 between fiber and matrix are performed. The domain is subjected to tensile loads, and the obtained stresses from PINNs are compared against that obtained from FE. Figure. \ref{fig:high_modulus_ratio} represents the $\sigma_{yy}$ stress for various modulus ratios.
\section{Surrogate modeling and uncertainty quantification}
\label{sec:uq}
\par To predict a wide range of futuristic structural behavior for a spectrum of variations, pertinent numerical models are necessary for all fields of engineering, and science \cite{8770759, betz2014numerical}. UQ helps quantify system errors and identifies the suitable parameters that are highly dynamic. The standard numerical approaches fail to generate reliable results in less time due to their heavy dependence on the meshing. If the geometry or material properties are changed, separate simulations need to be run to address the associated UQ. As pointed out in Section \ref{sec:intro}, PINNs can solve the problem of random differential equations where domains can have varying geometry or material properties. In the present study, the random governing equations can be described as
\begin{equation}
    \nabla.\boldsymbol{{\sigma}}(\boldsymbol{{x}},p_{g/m}) = 0,
    \label{Eq.stochastic governing eqs}
\end{equation}
\subsection{Randomness in geometry}
\label{sec:randomness_geometry}
\begin{figure}[H]%[!htbp]
\centering
  \begin{subfigure}[H]{0.49\textwidth}
     \includegraphics[width=\textwidth]{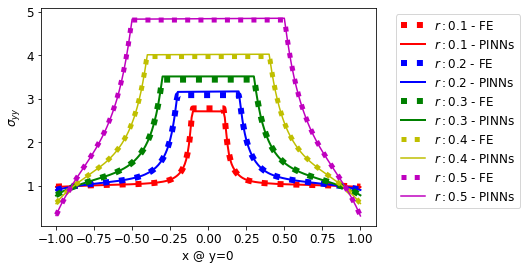}
     \caption{$\sigma_{yy}$}
     \label{fig:1}
  \end{subfigure}
  \begin{subfigure}[H]{0.49\textwidth}
     \includegraphics[width=\textwidth]{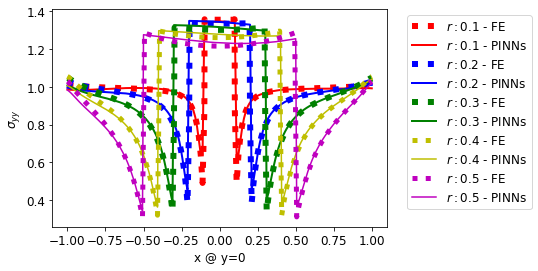}
     \caption{$\sigma_{yy}$}
     \label{fig:2}
  \end{subfigure}
  \caption{Parametric study for a single void and fiber with radius as a parameter. (a) Void, (b) fiber.}
  \label{fig:Parametric_study_for_single_void_and_inclusion_with_radius_as_a_parameter}
\end{figure}

The constitutive equation describing the stochastic nature of geometry with a single void / fiber is described as  
\begin{subequations}
\begin{align}
  \boldsymbol{\sigma}(\boldsymbol{x},p_g)=\boldsymbol{\mathcal{C}}:\epsilon(\boldsymbol{x},p_g) = 0 \hspace{2mm} \boldsymbol{{x}}\in\Omega,
  \label{Eq. varying radius constitutive equation}\\
\boldsymbol{\mathcal{C}}=\boldsymbol{\mathcal{C}_o},
\label{Eq. varying radius modulus}
\end{align}
\end{subequations}
The random parameter $p_g$ serves as the radius of the void / fiber within a given range. As shown in \eqn{Eq. varying radius modulus} the material property matrix $\boldsymbol{\mathcal{C}}$ is assumed to be constant. In addition to the spatial dimensions $x$ and $y$, $p_g$ acts as an extra input to the neural network. With one complete training of the neural network, any number of stress states can be inferred depending on the radius of the void. After sufficient training, five different stress states corresponding to the radii 0.1--0.5 are extracted from the PINNs solver. Figure. \ref{fig:Parametric_study_for_single_void_and_inclusion_with_radius_as_a_parameter} represents the $\sigma_{yy}$ stress comparison from PINNs and the FE solver. It is noteworthy that the results from PINNs based solver are only from one single training whereas, for FE solver, they are from five independent simulations. 

% \begin{figure}[H]%[!htbp]
% \centering
%      \includegraphics[width=0.5\textwidth]{figs/single_void_params/stresscomp_voids_all_sizes.png}
%      \caption{Parametrized linear momentum equation with radius}
%      \label{fig: Parametrized linear momentum equation with radius}
% \end{figure}

\begin{figure}[H]%[!htbp]
\centering
  \begin{subfigure}[H]{0.49\textwidth}
     \includegraphics[width=\textwidth]{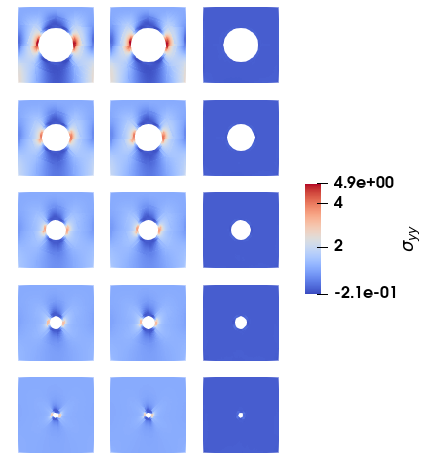}
     \caption{}
     \label{fig:1}
  \end{subfigure}
  \begin{subfigure}[H]{0.49\textwidth}
     \includegraphics[width=\textwidth]{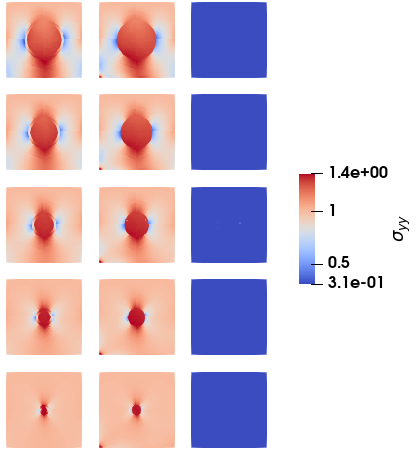}
     \caption{}
     \label{fig:2}
  \end{subfigure}
  \caption{Randomness in geometry. Each row contains FE, PINNs and their difference in order. (a) Varying void radius, (b) varying fiber radius.}
  \label{fig:Randomness_in_geometry_Each_row_contains_FE_Modulus_and_their_difference_in_order.}
\end{figure}
Figure. \ref{fig:Randomness_in_geometry_Each_row_contains_FE_Modulus_and_their_difference_in_order.} shows the comparison of stress fields obtained from PINNs and FE solver. This procedure is highly computationally efficient and leads to a robust design approach.

\subsubsection{Calculation of effective material properties}
Analytical solutions for effective material properties are only available for specific geometries of the fiber \cite{molkov1985effective,rodriguez2012two}. In order to take into account the effect of a variety of fiber shapes and their volume fractions, the immediate alternative is numerical approaches like FE. In this section, we demonstrate that PINNs can also be used to determine homogenized properties of FRCs within a reasonable tolerance. The averaged quantities of stresses $\sigma_{ij}$ and strains $\epsilon_{ij}$ are given by
\begin{subequations}
\begin{align}
    \langle{\sigma_{ij}}\rangle=\frac{1}{N}\mathlarger{\mathlarger{\mathlarger{\Sigma}}}_{c=1}^N \sigma_{ij}(c),
    \label{Eq. average stress}\\
    \langle{\epsilon_{ij}}\rangle=\frac{1}{N}\mathlarger{\mathlarger{\mathlarger{\Sigma}}}_{c=1}^N \epsilon_{ij}(c),
    \label{Eq. average strain}
\end{align}
\label{eq:average_stress_strain}
\end{subequations}
where $N$ represents the number of collocation points and $\sigma_{ij}(c)$ and $\epsilon_{ij}(c)$ represent stress and strain states at collocation point $c$. Transverse Young's modulus defined for loading in $y$ direction as shown in Figure. \ref{fig:Initial_setup_of_the_problem}, is calculated as the ratio of average stress to the average strain. Loading and boundary conditions are the same as shown in Figure. \ref{fig:Parametric_study_for_single_void_and_inclusion_with_radius_as_a_parameter}.
\begin{figure}[H]%[!htbp]
\centering
  \begin{subfigure}[H]{0.49\textwidth}
     \includegraphics[width=\textwidth]{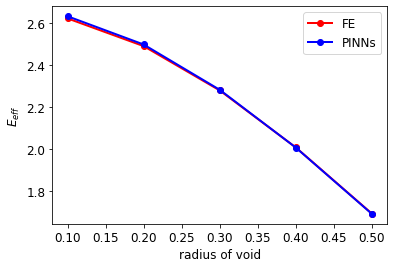}
     \caption{}
     \label{fig:effective_void}
  \end{subfigure}
  \begin{subfigure}[H]{0.49\textwidth}
     \includegraphics[width=\textwidth]{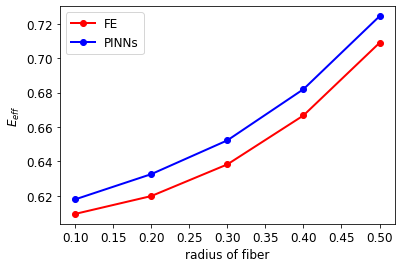}
     \caption{}
     \label{fig:effective_inclusion}
  \end{subfigure}
  \caption{Effective transverse elastic modulus with varying radius of (a) void, (b) fiber.}
  \label{fig: Effective material properties in transverse direction with varying radius}
\end{figure}
The evolution of the effective modulus (non dimensional) in the presence of varying radius is shown in Figure. \ref{fig: Effective material properties in transverse direction with varying radius} and some observations are as following:
\begin{itemize}
    \item Effective modulus of elasticity of the composite shows strong dependence on the volume fraction of the void \cite{watt1976elastic}. It is inferred from Figure. \ref{fig:effective_void} that the effective transverse modulus decreases rapidly as the size of the void increases.
    \item Effective modulus increases with the increase in radius of the fiber, which demonstrates that the effective modulus of the composite is driven by the volume fraction of the material with high modulus.
    \item It is evident from Figure. \ref{fig:effective_inclusion} that PINNs can capture the homogenized quantities with a maximum error tolerance of $2.3\%$, which is for 0.3 radius fiber.
\end{itemize}
\subsection{Randomness in material properties}
\begin{figure}[H]%[!htbp]
\centering
  \begin{subfigure}[H]{0.495\textwidth}
  \centering
     \includegraphics[width=0.8\linewidth,height=0.6\linewidth]{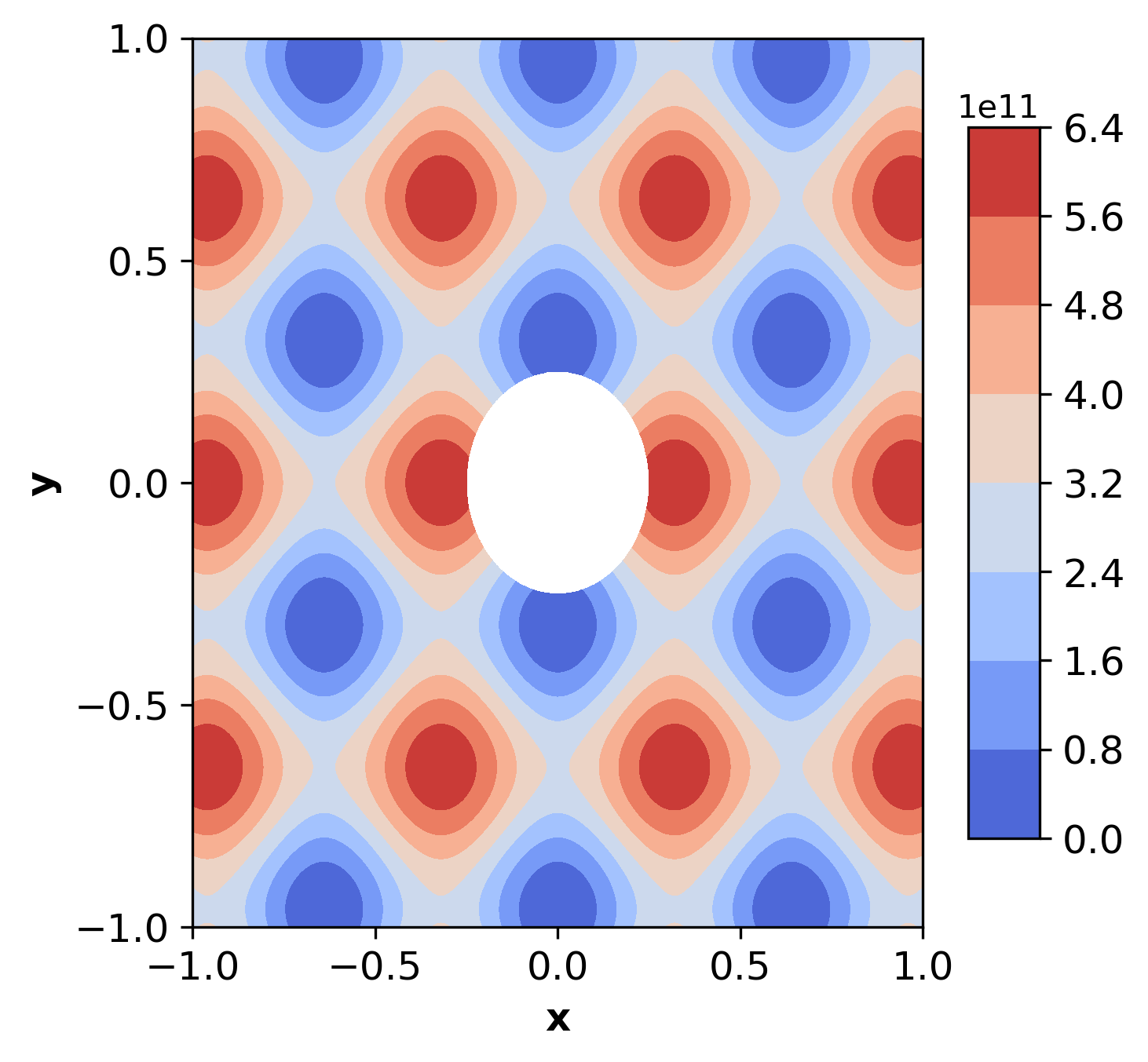}
     \caption{}
     \label{fig:1}
  \end{subfigure}
  \begin{subfigure}[H]{0.495\textwidth}
  \centering
     \includegraphics[width=0.8\linewidth,height=0.6\linewidth]{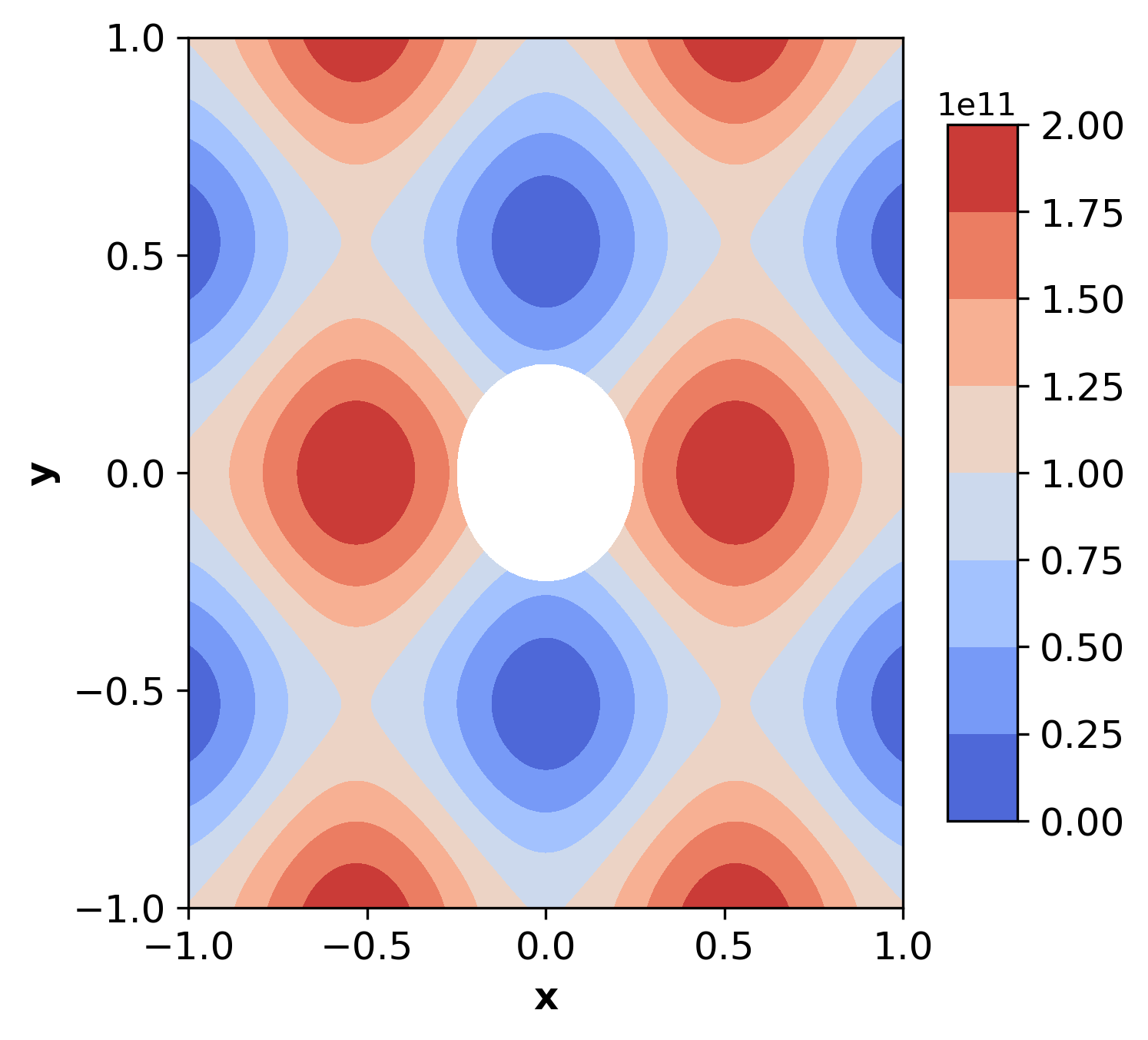}
     \caption{}
     \label{fig:2}
  \end{subfigure}
   \begin{subfigure}[H]{0.495\textwidth}
   \centering
     \includegraphics[width=0.8\linewidth,height=0.6\linewidth]{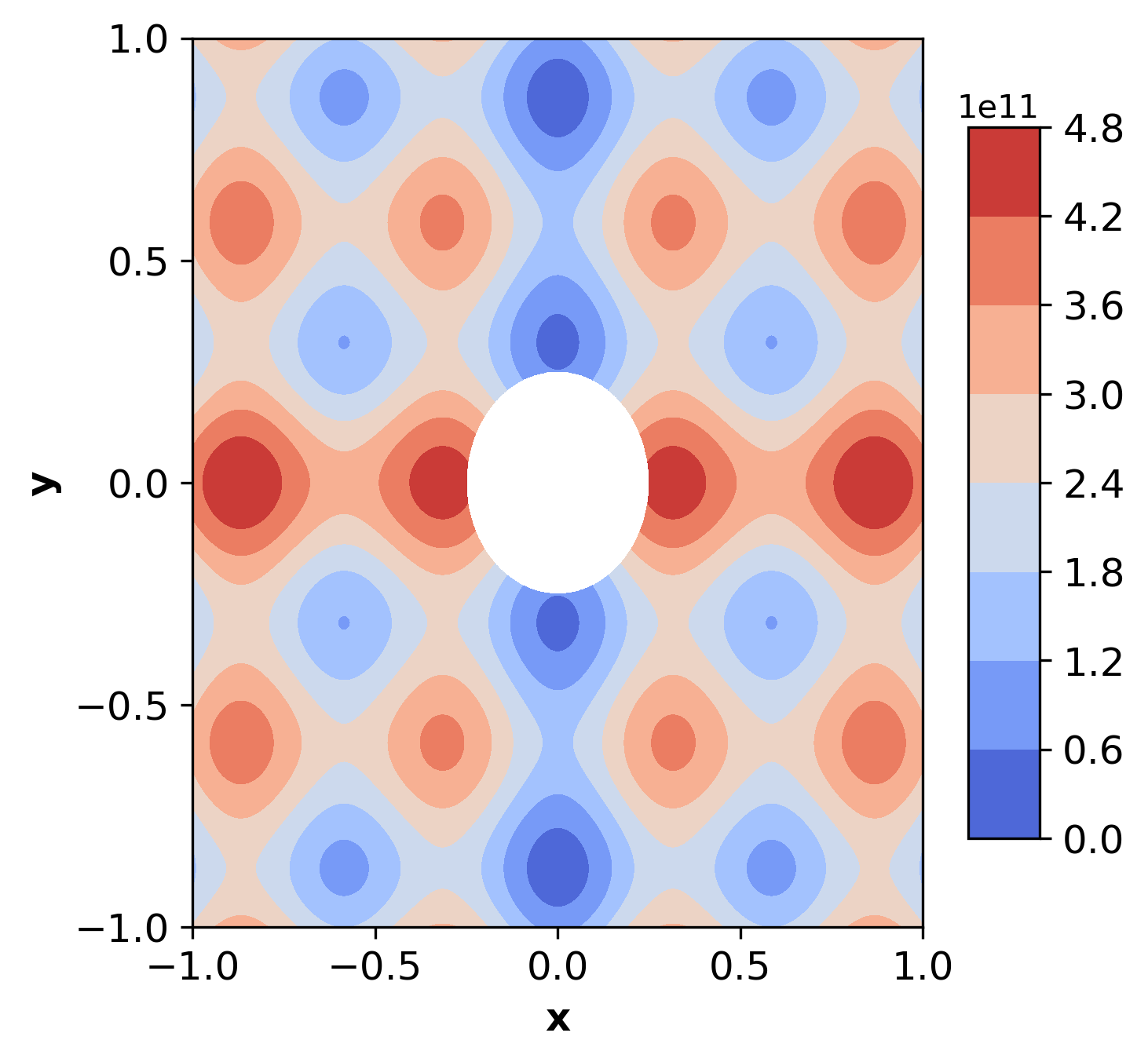}
     \caption{}
     \label{fig:1}
  \end{subfigure}
  \begin{subfigure}[H]{0.495\textwidth}
  \centering
     \includegraphics[width=0.8\linewidth,height=0.6\linewidth]{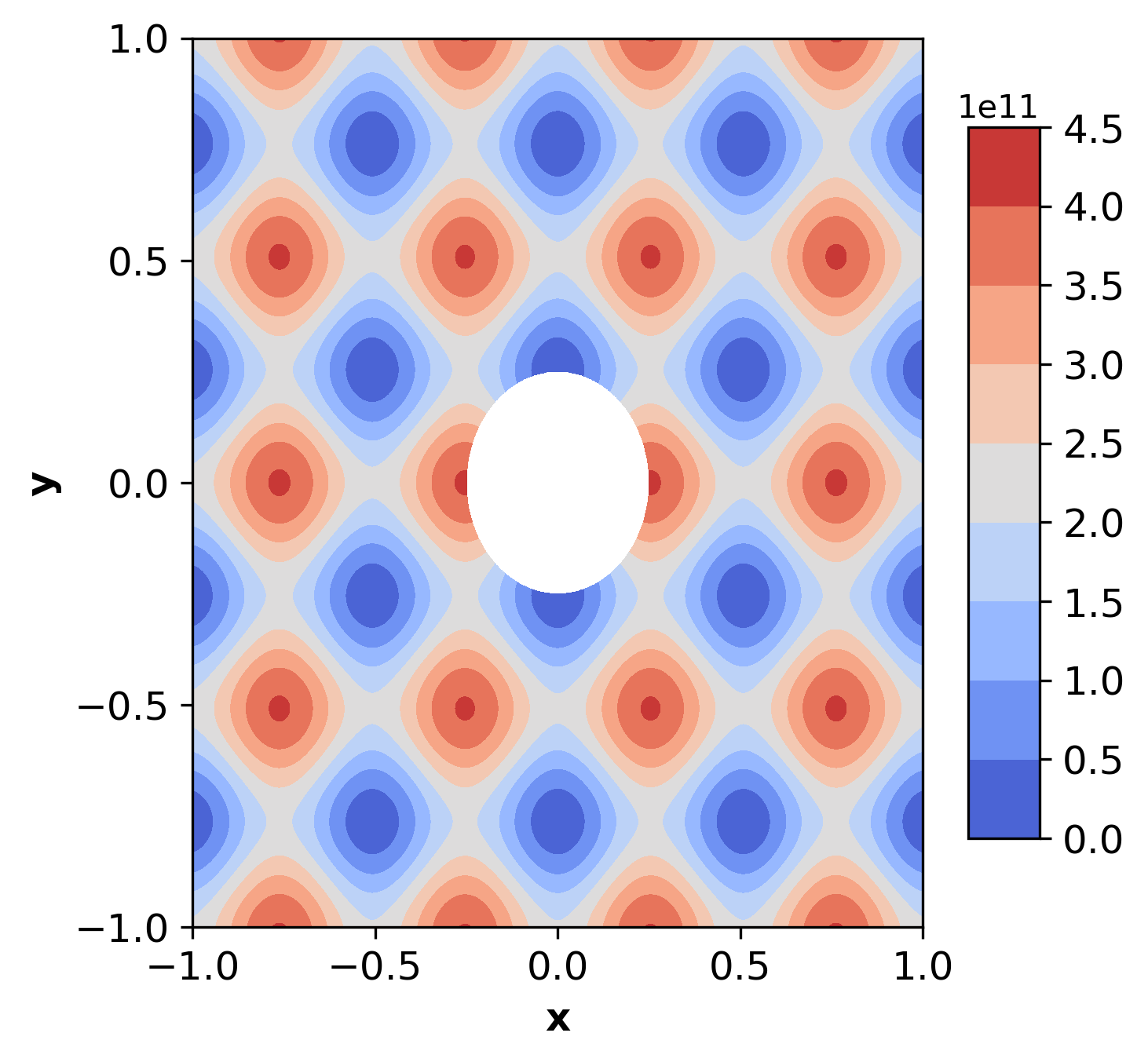}
     \caption{}
     \label{fig:2}
  \end{subfigure}
  \caption{Variation of Young's modulus with probablistic parameter. (a) Mode 1, (b) Mode 2, (c) Mode 3, (d) Mode 4.}
  \label{fig:stiffness_modes}
\end{figure}

The material properties in heterogeneous materials are substantially different from one location to another. In the present section, a smooth random field is assumed for the elasticity tensor as a function of space in $\mathbb{R}^2$ as given by
\begin{subequations}
\begin{align}
\boldsymbol{\sigma}(\boldsymbol{x,p_m})=\boldsymbol{\mathcal{C}}(\boldsymbol{x,p_m}):\epsilon(\boldsymbol{x,p_m}),\\
\boldsymbol{\mathcal{C}}(\boldsymbol{x,p_m})=\boldsymbol{\mathcal{C}}_o\scalebox{2}{${\Sigma}$}_{i=1}^{n} (\sin(2\pi x {p_m}_i)^2+\cos(2\pi y {p_m}_i)^2)\times {p_m}_i,
\label{Eq.stiffness_random_field}
\end{align}
\end{subequations}
where $\boldsymbol{p_m}=\{{p_m}_1,{p_m}_2,...{p_m}_n\}$ is an $n$ dimensional vector that controls the magnitude of stiffness and $\boldsymbol{\mathcal{C}}_o$ represents the base stiffness and is chosen to be the same as $E_f$.
%Material properties are shown in Table \ref{table:parameters list}.
In the present study $n=2$ is considered i.e., the function describing elasticity tensor contains two independent parameters ${p_m}_1$ and ${p_m}_2$. These two parameters are additional inputs to the neural network solver and spatial coordinates, which increase the dimensions of the inputs. Various modes corresponding to the assumed function are shown in Figure. \ref{fig:stiffness_modes}.

\begin{figure}[H]%[!htbp]
\centering
  \begin{subfigure}[H]{0.5\textwidth}
     \includegraphics[scale=0.6]{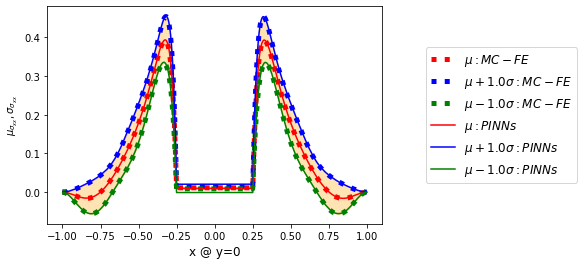}
     \caption{}
     \label{fig:1}
  \end{subfigure}
  \begin{subfigure}[H]{0.5\textwidth}
     \includegraphics[scale=0.6]{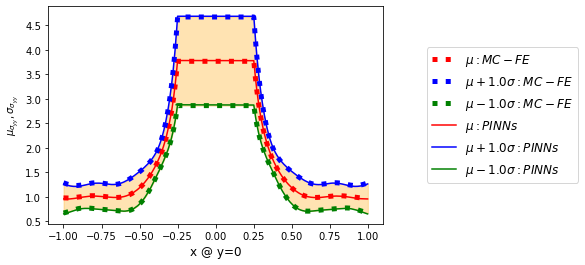}
     \caption{}
     \label{fig:2}
  \end{subfigure}
  \caption{Mean and standard deviation for MC--FE and PINNs methods compared for (a) $\sigma_{xx}$, (b) $\sigma_{yy}$.}
  \label{fig:comparison_between_Monte_carlo_FE_and_PINNs}
\end{figure}

\begin{figure}[H]%[!htbp]
\centering
  \begin{subfigure}[H]{0.48\textwidth}
     \includegraphics[scale=0.5]{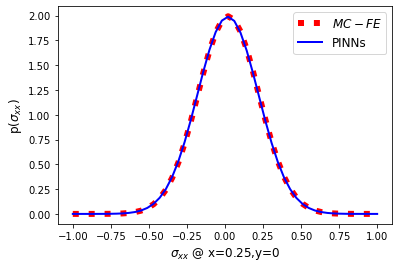}
     \caption{}
     \label{fig:1}
  \end{subfigure}
  \begin{subfigure}[H]{0.48\textwidth}
     \includegraphics[scale=0.5]{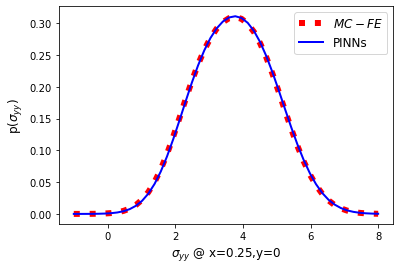}
     \caption{}
     \label{fig:2}
  \end{subfigure}
  \caption{Probability density functions for (a) $\sigma_{xx}$ and (b) $\sigma_{yy}$.}
  \label{fig:Probability_density_functions_for_sigma_xx_and_sigma_yy}
\end{figure}

In order to address the stiffness variation across the domain, each finite element simulation requires more than $1 \times 10^6$ elements, making the Monte Carlo simulations very intensive and time--consuming. The simulations are conducted using plane strain elements with reduced integration. 1000 realizations for Monte Carlo Finite element simulations are set up. Similarly, from the converged solution of PINNs, 1000 realizations are extracted for randomly chosen parameters of $\boldsymbol{p_m}$. The $\sigma_{xx}$ and $\sigma_{yy}$ stresses are extracted, and mean, and standard deviation from both PINNs and Monte Carlo FE (MC--FE) are compared. The confidence interval from both approaches is shown in Figure. \ref{fig:comparison_between_Monte_carlo_FE_and_PINNs} and are in excellent agreement. Additionally, Figure. \ref{fig:Probability_density_functions_for_sigma_xx_and_sigma_yy} shows the probability density functions for $\sigma_{xx}$ and $\sigma_{yy}$ at $x=0.25,y=0$ location and an excellent agreement is observed between PINNs and MC-FE methodologies.

\section{Conclusions}
\label{sec:conclusions}
In this work, the application of PINNs has been demonstrated to address the elastic deformation behavior of heterogeneous materials, similar to Fiber Reinforced Composites (FRCs), containing voids and fibers. A PINN based model has been developed for solving PDEs representing the balance of momentum. Material properties representing linear isotropic elasticity are considered. Our PINN model has been implemented in the Modulus framework named Modulus developed by NVIDIA. We explore the heterogeneity due to fibers and voids that are highly discontinuous at interfaces due to sudden changes in elasticity constants. Such changes induce high gradients in their stress values at the interface regions, which may later manifest as the sources of cracks etc. The key conclusions are as follows:
\begin{itemize}
\item The PINNs are able to capture discontinuities precisely, and the stress fields are shown to compare very well against the solutions from commercial FE solver.
\item Numerical experiments are conducted for single fibers / voids of different shapes and fiber--reinforced composites consisting of high modulus regions (fibers) and low modulus regions (voids) under tensile and shear loading. We find that PINNs can capture the stress states correctly even when the modulus ratio is as high as 100.
\item PINNs can be effectively used for determination of effective properties for FRC with complicated designs.
\item PINNs can be employed for surrogate modeling and uncertainty quantification (UQ). The results from PINNs for the responses for variations in material properties and composite designs show an excellent agreement against the responses from MC coupled with FE.
\end{itemize}
In this work it is echoed that PINNs can be used as a highthroughput and meshless surrogate model for designing novel composites. UQ is an added advantage of it. This work opens a few interesting avenues and challenges to be taken up by us. Due to the inherent capabilities of inverse design using PINNs, matrix and inclusion material properties can be estimated efficiently. It can be also used for topology design and optimization for meta--materials.
\section*{Acknowledgement}
AA acknowledges the financial support provided by IIT Bombay for PhD student BVSSB. AA also acknowledges the partial financial support from Aeronautics Research Development Board (ARDB) vide Grant \# GTMAP1952 for part of a Computational Facility that was used for this work. AA and BVSSB acknowledge discussions with Ritesh Dadhich about GPUs, PDEs and logistics.
%% If you have bibdatabase file and want bibtex to generate the
%% bibitems, please use
%%
%\clearpage
\bibliographystyle{elsarticle-num} 
\bibliography{cas-refs}

%% else use the following coding to input the bibitems directly in the
%% TeX file.

% \begin{thebibliography}{00}

% %% \bibitem{label}
% %% Text of bibliographic item

% \bibitem{}

% \end{thebibliography}
\end{document}